# Elastic Properties of Chemical-Vapor-Deposited Monolayer MoS$_2$, WS$_2$, and Their Bilayer Heterostructures

*Kai Liu,[†,‡] Qimin Yan,[#,∥] Michelle Chen,[†] Wen Fan,[†,] Yinghui Sun,[∥] Joonki Suh,[†] Deyi Fu,[†] Sangwook Lee,[†] Jian Zhou,[†] Sefaattin Tongay,[†] Jie Ji, Jeffrey B. Neaton,[#,∥] Junqiao Wu[†,‡,\*]*

[†] Department of Materials Science and Engineering, University of California, Berkeley, California 94720, United States

[‡] Materials Sciences Division, Lawrence Berkeley National Laboratory, Berkeley, California 94720, United States

[#] Molecular Foundry, Lawrence Berkeley National Laboratory, Berkeley, California 94720, United States

[∥] Department of Physics, University of California, Berkeley, California 94720, United States

Department of Thermal Science and Energy Engineering, University of Science and Technology of China, Anhui 230027, China

ABSTRACT

Elastic properties of materials are an important factor in their integration in applications. Chemical vapor deposited (CVD) monolayer semiconductors are proposed as key components in industrial-scale flexible devices and building blocks of 2D van der Waals heterostructures. However, their mechanical and elastic properties have not been fully characterized. Here we report high 2D elastic moduli of CVD monolayer $MoS_2$ and $WS_2$ (~ 170 N/m), which is very close to the value of exfoliated $MoS_2$ monolayers and almost half the value of the strongest material, graphene. The 2D moduli of their bilayer heterostructures are lower than the sum of 2D modulus of each layer, but comparable to the corresponding bilayer homostructure, implying similar interactions between the hetero monolayers as between homo monolayers. These results not only provide deep insight to understanding interlayer interactions in 2D van der Waals structures, but also potentially allow engineering of their elastic properties as desired.

Two-dimensional (2D) semiconducting transition metal dichalcogenides (TMDs), such as MoS$_2$ and WS$_2$,[1,2] receive growing attention owing to their bandgap crossover from indirect in the bulk to direct in the monolayer,[3-6] and a range of potential applications in optoelectronic and photonic devices.[7-9] Chemical vapor deposition (CVD) has been developed to synthesize low-cost and scalable 2D TMD monolayers for practical device applications.[10-14] It is also proposed as the most feasible approach to fabricate 2D heterostructures at industrial scale by simply stacking CVD-grown monolayers.[15] The diversity of 2D crystals results in a large number of possible 2D heterostructures that possess interesting charge-splitting functions for applications.[15-19] However, in contrast to their extensively studied electrical and optical properties, elastic and mechanical properties of 2D TMDs and their heterostructures have not been well characterized.

Elastic modulus is a basic parameter to determine mechanical properties of materials, and is of vital importance in recent applications of flexible and stretchable electronics and photonics.[20] 2D crystals have already been employed as key components in flexible devices due to their atomic thickness and ultrahigh flexibility.[21-24] However, reports on elastic properties of 2D TMDs are limited to less-defective, exfoliated MoS$_2$ with scattered experimental results,[25-27] while CVD-grown 2D TMDs have not been measured. On the other hand, interlayer coupling of 2D heterostructures plays a great role in the performance of devices. Although the coupling has been investigated electrically and optically in various 2D heterostructures,[15-19] a mechanical probing of the interlayer coupling is complementary but currently lacking.

In this work we measured the elastic modulus of CVD-grown monolayer MoS$_2$ and WS$_2$, and probed the interlayer interaction of their heterostructures. The CVD-grown MoS$_2$ and WS$_2$ are found to have similarly high 2D elastic modulus, ~ 170 N/m, very close to the value of exfoliated MoS$_2$, and almost half of that of graphene. Theoretical simulations confirm that MoS$_2$ and WS$_2$ have nearly the same lattice constants and elastic properties. The 2D moduli of heterostructures are slightly lower than the sum of 2D modulus of each layer, but comparable to the corresponding bilayer homo structures, implying similar interactions between hetero monolayers compared to between homo monolayers. The interlayer coupling of different bilayer homo or hetero structures is also qualitatively compared. These results provide useful insight to understanding interlayer interactions in 2D materials, and their utilization in flexible devices.

Monolayer $MoS_2$ ($WS_2$) were synthesized by CVD on $SiO_2$ (300 nm thick)/Si substrates with solid $MoO_3$ ($WO_3$) and S as precursors, similar to the method published previously.[14] The as-grown samples (Fig. 1a) show isolated triangles of monolayer crystals at positions on the substrate slightly far away from the precursors, while closer to the precursors these triangles merge into a continuous film. The average size of these triangles is 10 ~15 µm. A thicker-layer island usually sits at the center of each triangle (Fig. 1b), which appears bigger and more evident in the continuous part (Fig. 1c); this is probably due to the nucleation of additional layers over the bottom monolayer. The as-grown $MoS_2$ or $WS_2$ monolayers were then transferred onto a holey $SiO_2$/Si substrate by a polydimethylsiloxane (PDMS) stamping process (Fig. S1 and S2). Figure 1d shows atomic force microscopy (AFM) image of a single $MoS_2$ monolayer triangle transferred onto the holey substrate. Photoluminescence (PL) mapping reveals a uniform light emission intensity from the $MoS_2$ monolayer in contact with the underneath $SiO_2$/Si substrate, while that of the suspended part over the holes is much brighter due to higher quantum yield (Fig. 1e), consistent with the report on exfoliated monolayer $MoS_2$.[3]

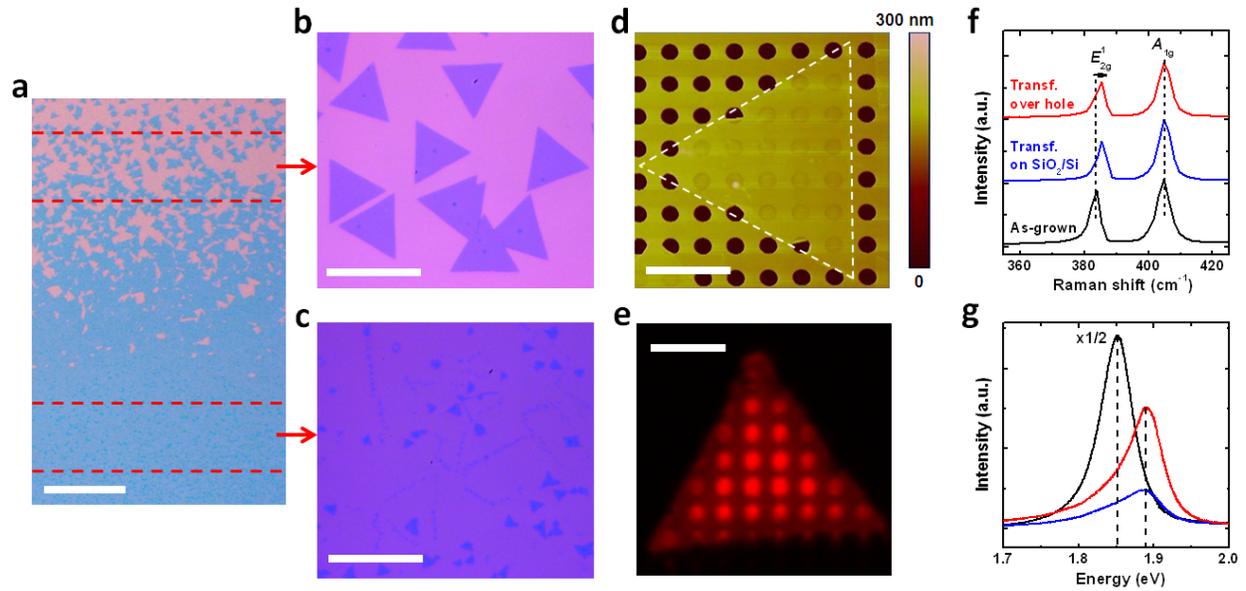

Figure 1. As-grown and transferred MoS$_2$ monolayer. (a) MoS$_2$ monolayer as grown on substrate, consisting of (b) isolated triangles and (c) continuous film of monolayers. (d) AFM topography of a triangle monolayer transferred onto the holey substrate. (e) Map of PL peak intensity from a triangle monolayer sitting on the holey substrate. (f, g) Raman and PL spectra of as-grown MoS$_2$, transferred MoS$_2$ in contact with SiO$_2$/Si, and suspended MoS$_2$ over holes. Scale bars, (a) 100 µm; (b and c) 10 µm; (d and e) 5 µm.

For 2D crystals, Raman and PL spectra are typically utilized to distinguish the number of layers[28, 29] as well as other effects such as strain and charge transfer.[30, 31] As shown in Fig. 1f, the out-of-plane Raman mode $A_{1g}$ remains at ~ 405 cm$^{-1}$, but the in-plane mode $E_{2g}^1$ blue-shifts from 384 cm$^{-1}$ in as-grown MoS$_2$ to ~ 386 cm$^{-1}$ in transferred MoS$_2$. The separation between the $E_{2g}^1$ and $A_{1g}$ modes after the transfer is 19 cm$^{-1}$, similar to that of strain-free exfoliated monolayer MoS$_2$ (typically < 20 cm$^{-1}$).[28, 29] The PL of the as-grown sample shows a strong peak centered at ~

1.85 eV, resulting from the A direct excitonic transition.[4] After the transfer, the PL intensity is dramatically reduced, and the PL peak is blue shifted to 1.89 eV (Fig. 1g). The transfer of CVD WS$_2$ monolayers shows similar results (Fig. S4). Both shifts of the PL and the in-plane Raman mode ($E_{2g}^1$) of the MoS$_2$ monolayer after the transfer can be attributed to the strain effect. It has been reported that a tensile strain softens PL and Raman modes of MoS$_2$ monolayer.[30] A similar strain may exist in as-grown MoS$_2$ monolayers due to the different thermal expansion coefficients of the materials in growth, and it will be released after the transfer such that the monolayer MoS$_2$ transferred is free of strain on the target substrate. The change of PL intensity may originate from the desorption / adsorption of molecules during the transfer.[31]

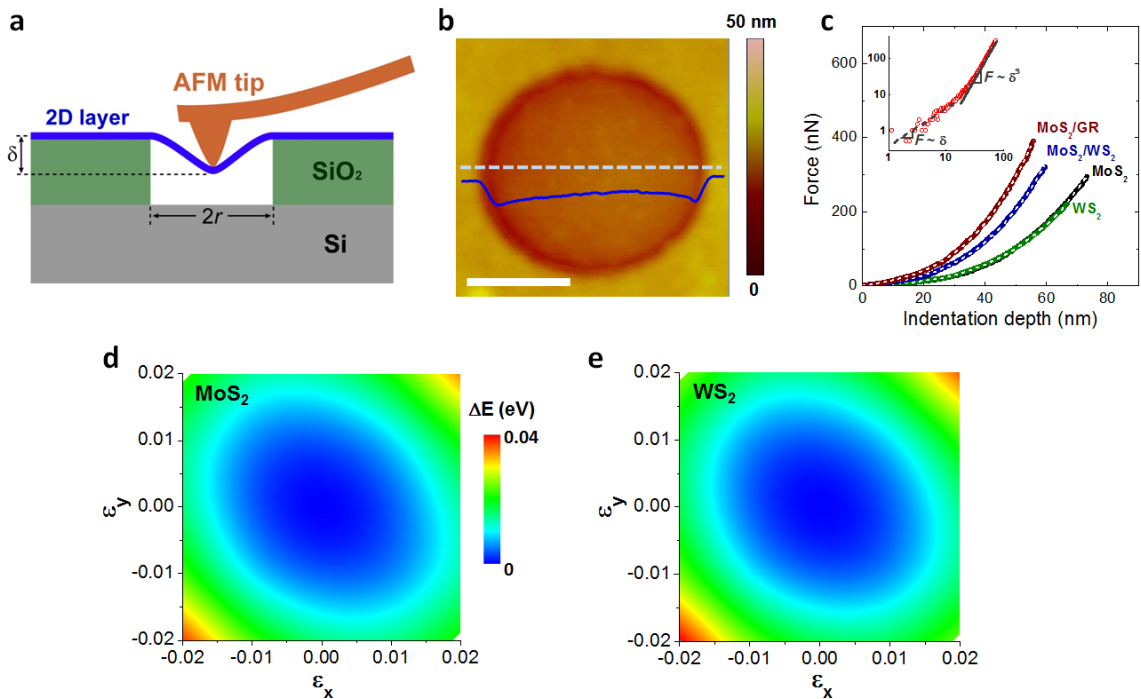

Figure 2. Elastic properties of 2D monolayers and heterostructures. (a) Illustration of the indentation measurement. (b) Typical AFM topography of a MoS$_2$ monolayer over a hole. Scale bar, 500 nm. (c) Force-displacement curves of different CVD monolayers and heterostructures.

The white dashed lines are fitted by the equation 1. (d and e) Contour plot of calculated elastic energy change for MoS$_2$ and WS$_2$ monolayers, respectively, under various biaxial strains.

The elastic moduli of 2D layers are measured by indenting the center of the suspended part as a circular membrane with an AFM tip (Fig. 2a). The tip is coated by diamond-like carbon and its diameter is ~ 20 nm. Tapping-mode AFM image displays that the 2D monolayer membranes are taut over the hole (Fig. 2b). AFM images of bilayer homo- or hetero-structures (Fig. S5-S7) do not show evidence of bubbles or wrinkles either, whether on the substrate or over holes. This benefits from the dry PDMS stamping process, a technique proven capable of avoiding ripples or wrinkles that would introduce errors in measuring the modulus of CVD graphene.[32] For 2D materials, the strain energy is normalized by the sheet area, giving rise to 2D stress $\sigma^{2D}$ and elastic modulus $E^{2D}$. MoS$_2$ and WS$_2$ have three-fold rotation symmetry, and are thus isotropic in plane. For such an ultrathin monolayer membrane clamped across a hole and indented at the center by a tiny tip ($r_{tip} \ll r_{hole}$), the bending modulus is negligible. The load is balanced by the pretension of the membrane and scales linearly with vertical deflection (F ~ δ) under small loads.[33] When the load is large, it is dominated by the stiffness of the membrane with a cubic relationship, F ~ δ$^3$ (inset of Fig. 2c).[34] The force-displacement relationship can be described approximately as[35]

$$F = (\sigma_0^{2D}\pi)\delta + (E^{2D}\frac{q^3}{r^2})\delta^3, \qquad (1)$$

where $F$ is the applied point load, $\delta$ is the indentation depth at the center of the membrane, $\sigma_0^{2D}$ is the pretension, $r$ is the radius of the hole, and $q$ is a dimensionless constant determined by the Poisson's ratio ($\nu$) of the membrane, obeying $q=1/(1.05 - 0.15\nu - 0.16\nu^2)$. We utilized the values from first-principles calculations (details below), which are $\nu$=0.25 and $\nu$=0.22 for MoS$_2$ and WS$_2$, respectively, because of discrepancy in Poisson's ratios of MoS$_2$ in previous studies[25-27] and the lack of experimental value for WS$_2$ in literature. For graphene, $\nu$=0.165.[35] $\sigma_0^{2D}$ and $E^{2D}$ can be derived by least-square fitting of the experimental force-displacement data with Eq. 1 (Fig. 2c).

To explore the elastic properties of monolayer MoS$_2$ and WS$_2$, we first performed first-principles calculation using the plane-wave projector augmented wave (PAW) method as implemented in the VASP code. The exchange correlation potential is approximated by the GGA-PBE method.[36, 37] Details of calculation can be found in the Supporting Information Fig. S8 and the experimental section. Figure 2d and 2e show the contour plot of change in elastic energy as a function of strain along the x and y directions. The 2D elastic moduli and Poisson's ratios derived from the calculated data are 123 N/m and 0.25 for MoS$_2$, and 137 N/m and 0.22 for WS$_2$, respectively. The calculated Poisson's ratio of MoS$_2$ is close to the experimental value (~0.27 in ref [38]), while the calculated value of WS$_2$ provides a reference for our experimental fitting of modulus as it has never been measured.

We measured 7~15 suspended circular membranes for each type of 2D layers (graphene, MoS$_2$, WS$_2$, and their heterostructures), and indented each membrane with 2~3 different applied forces. For bilayer heterostructures, $q$ is taken as the average value of the two layers, because of the weak interlayer van der Waals interaction and the resultant difficulty in determining the Poisson's ratios of heterostructures by the calculation. On the other hand, the separate $q$ values of MoS$_2$, WS$_2$, and graphene are very close to each other (0.998, 0.991, 0.980, respectively); therefore, an average value of $q$ will not introduce significant errors for the fitting.

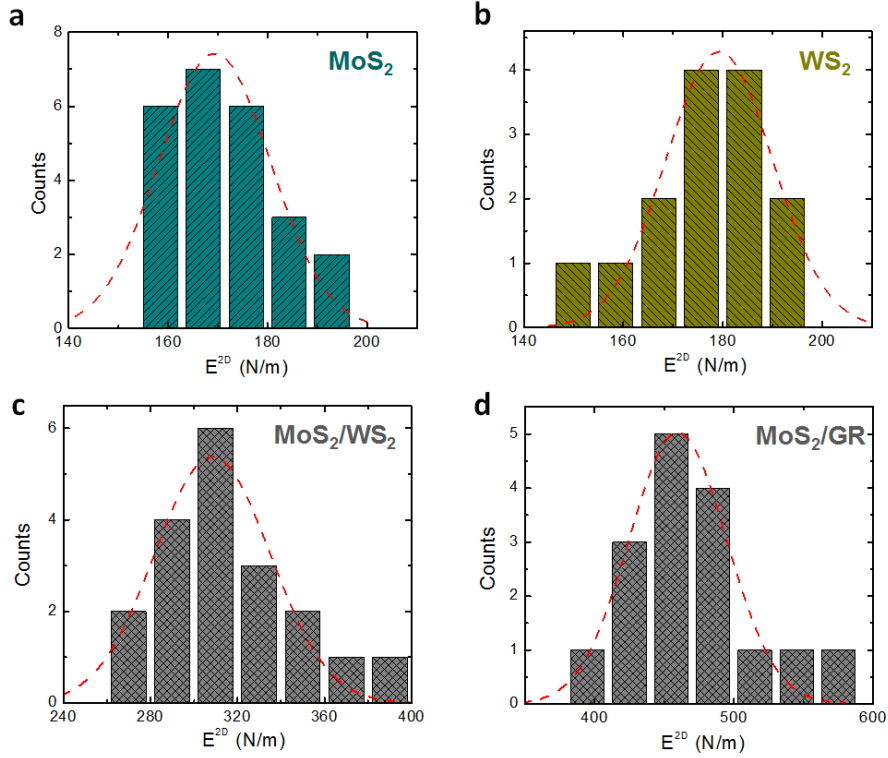

Figure 3. Histogram of $E^{2D}$ for different 2D layers and heterostructures. (a) Triangles of CVD MoS$_2$. (b) CVD WS$_2$. (c) MoS$_2$/WS$_2$ heterostructure. (d) CVD MoS$_2$/exfoliated graphene heterostructure.

Figure 3 shows the statistical histogram of the 2D modulus of MoS$_2$, WS$_2$ monolayers, and their heterostructures, MoS$_2$/WS$_2$ and MoS$_2$/graphene. The 2D moduli are 171± 11 N/m and 177± 12 N/m for MoS$_2$ and WS$_2$ (Fig. 3a and 3b), respectively, where the values before ± are the average values and the errors are standard deviations of the measurements. These values are nearly half the value of graphene (349 ± 12 N/m in our experiment and ~340 N/m in literatures), the strongest 2D material.[32, 35] The small difference in the modulus between MoS$_2$ and WS$_2$ is consistent with the calculation, although both experimental values are higher than the calculated values. This observation is consistent with the fact that the GGA approximation usually underestimates the bulk modulus of many traditional semiconductor materials.[39]

2D bilayer heterostructures were prepared by stacking different 2D layers in sequence (Fig. S5 and S7). The moduli of $MoS_2/WS_2$ and $MoS_2$/graphene heterostructures are measured to be 314± 31 N/m and 467± 48 N/m (Fig. 3c and 3d), respectively, which are both lower than the summed modulus of the consisting layers (348 N/m and 520 N/m, respectively). Figure 4a shows all the experimental data of $\sigma_0^{2D}$ and $E^{2D}$ for different 2D layers and heterostructures. The pretension of 2D layers depends not only on the transfer process, but also on their intrinsic mechanical properties, because this parameter relates to the elastic energy of pretension after the transfer. Therefore, different monolayers or heterostructures can have different pretension values. The average pretensions are 0.11 ± 0.04 N/m, 0.15 ± 0.03 N/m, and 0.20 ± 0.05 N/m for monolayer $MoS_2$, $WS_2$ and graphene, respectively. The positive pretensions may originate from adhesion of the monolayer membrane to the sidewall of the hole (Fig. 2b), also consistent with the reports on exfoliated graphene and $MoS_2$.[25, 35] Interestingly, the pretensions of the heterostructures, $MoS_2/WS_2$ and $MoS_2$/graphene, are almost the summed pretension of both layers, which are 0.25 ± 0.05 N/m and 0.35 ± 0.05 N/m, respectively. This indicates that the pretension is simply accumulated as the layers are stacked sequentially by the same transfer process.

The simple stacking process for fabricating heterostructures has a general concern that whether the van der Waals heterostructures have strong interlayer interaction. While optical and electrical methods investigate the electronic coupling that has a strong influence on band structure renormalization,[16, 17] mechanical measurements can also probe the coupling force between the layers. In our experiments, none of the force-displacement curves of monolayers shows evidence of sliding (*e.g.*, irreversible force-displacement dependence). Therefore, the bottom layer in direct contact with the substrate should be firmly clamped onto the substrate within the range of load applied. In the extreme case of very strong interlayer interactions in a bilayer structure, there is no interlayer sliding allowed, and both layers contribute to the 2D modulus measured with our method. In the opposite extreme, *i.e.*, in the absence of interaction between layers, the top layer is free to slide against the bottom layer, and the measured modulus of the bilayer is solely given by the bottom layer that is clamped by the substrate. Thus, the measured modulus of a 2D bilayer hetero- or homo-structure can be phenomenologically described by

$$E^{2D}_{measured} = E^{2D}_{bottom} + \alpha E^{2D}_{top}, \qquad (2)$$

where $E^{2D}_{measured}$, $E^{2D}_{bottom}$ and $E^{2D}_{top}$ are the 2D modulus of, respectively, the bilayer, the bottom layer, and the top layer, and the "interaction coefficient" $\alpha$ ranges from 0 to 1 describing the contribution of the top layer to the measured value. $\alpha$ depends on the interlayer friction coefficient, the van der Waals interaction between layers, and the strain that relates to the indentation depth. Thus, this factor is not independent of strain. In our experiments, however, the average indentation depth is controlled at a very small range (53~61 nm). Thus the strain applied on all of bilayer structures is nearly the same. Therefore, $\alpha$ can be used to compare the interlayer coupling in different bilayer homo- or hetero-structures.

Besides the nanoindentation, nanoscale shearing or telescopic sliding can also be used to measure the interlayer interaction in graphite[40] or in multiwalled nanotubes.[41, 42] Compared to the nanoindentation, these methods provide a more direct way to quantify the interlayer friction. However, the nanoindentation is a relatively simple method to qualitatively compare the interlayer coupling in different bilayer structures. The lower effective modulus of $MoS_2/WS_2$ heterostructures compared to the consisting layers gives $\alpha = 0.80$, which indicates that a slight interlayer sliding probably occurred between the $MoS_2$ and $WS_2$ layers during the measurement. We note that the 2D modulus of exfoliated bilayer $MoS_2$ was also reported to be lower than twice the value of monolayer.[25] As a comparison, we also measured the modulus of exfoliated bilayer $MoS_2$ (Fig. S6), which is $300 \pm 13$ N/m with $\alpha = 0.75$ (Fig. 4b). Since $MoS_2$ and $WS_2$ used in our experiments have nearly the same 2D modulus, this result suggests that the interaction between the $MoS_2$ and $WS_2$ layer is comparable to that in the bilayer $MoS_2$.

Among the various 2D van der Waals heterostructures, graphene/TMD has recently attracted much interest.[16-18] We have formed graphene/$MoS_2$ heterostructures by exfoliating monolayer graphene onto a holey substrate followed by stamping CVD $MoS_2$ (Supporting Information, Fig. S7). We note that the 2D modulus of graphene depends very linearly on the number of layers (data not shown here), indicating a relatively strong interaction between graphene homolayers preventing interlayer sliding during the measurement, corresponding to an $\alpha$ value of almost 1 (Fig. 4b). The $MoS_2$/graphene heterostructure, however, only gives rise to $\alpha = 0.69$ (Fig. 4b), indicating a moderately strong interaction.

In these atomic structures, 2D modulus is more intrinsic. However, the conversion of 2D to 3D modulus enables a comparison of the modulus of 2D layers with the conventional Young's modulus of bulk materials. For the layered materials, the bulk can be considered as the stacking of a large number of monolayers. The thickness of the bulk is determined by the number of the layers multiplied by the interlayer distance. Therefore, one needs to divide the 2D value of the monolayer by the interlayer distance in order to convert into the normal 3D Young's modulus and compare to each other. This is also the method used in literature to calculate the 3D modulus of graphene[35] and exfoliated monolayer $MoS_2$.[25] Figure 4c summarizes the Young's moduli of three types of 2D materials ($MoS_2$, $WS_2$, and graphene) obtained in this work as well as those reported in literature. The interlayer distance used is 0.34 nm for graphene and 0.65 nm for both $MoS_2$ and $WS_2$. The corresponding 3D modulus is $264 \pm 18$ GPa for $MoS_2$, $272 \pm 18$ GPa for $WS_2$, and $1025 \pm 35$ GPa for graphene. The modulus of CVD $MoS_2$ is consistent with the result of exfoliated $MoS_2$ monolayer (~ 270 GPa).[25] Both moduli of monolayer $MoS_2$ and $WS_2$ are higher than experimental values of bulk $MoS_2$ (~240GPa)[38] and multilayer $WS_2$ nanotubes (~ 170 GPa),[43] which may suggest a dependence of the Young's modulus on the number of layers due to the interlayer sliding in multilayers or bulk.

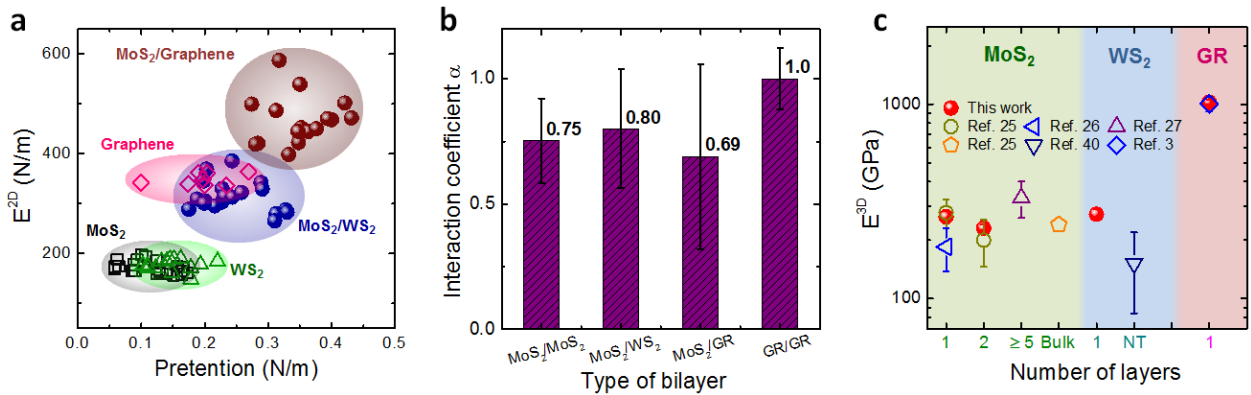

Figure 4. Comparison of elastic properties of different 2D layers. (a) Experimental data of 2D modulus and pretension for various 2D layers and heterostructures. (b) Interaction coefficients for different types of bilayers. (c) Comparison of Young's modulus of 2D monolayers with multilayers and bulk in the literature.

In summary, we experimentally and theoretically investigated the elastic properties of CVD-grown $MoS_2$, $WS_2$ and their heterostructures. Our work not only quantifies the elastic modulus of these 2D structures, but also defines an interaction coefficient between neighboring monolayers that measures the interlayer coupling. As interesting devices can be built from CVD 2D semiconductors and their heterostructures, our results provide calibrated values of their elastic modulus for various applications, especially in flexible electronic and mechanical devices. The studies of interlayer coupling provide a way to probe and understand electronic and mechanical coupling between monolayers in a variety of 2D structures.

**Methods**

**Synthesis of $MoS_2$ and $WS_2$.** $MoS_2$ and $WS_2$ monolayers were grown onto $SiO_2$/Si substrates using an ambient vapor transport technique. Prior to the growth, $SiO_2$/Si substrates were treated with Piranha solution for 2 hours. Our results show that this step is critical to achieve clean, large single crystal domain, and continuous area monolayers. After the piranha treatment, the samples were washed with DI water and dried using $N_2$ gas. 3 mg of $MoO_3$ powder is loaded in the alumina crucible and the samples are placed facing down. $MoO_3$ source was placed down-stream from the sulfur source at ~19 cm away from the sulfur containing crucible. The system was purged with ultra-high purity $N_2$ gas with 500 sccm flow rate for 10 minutes and heated up to 300 °C for 10 minutes with 100 sccm. Sulfur source melted at 600°C and $S_2$ gas was delivered to the growth zone at 2 sccm flow rate. The growth temperature was 690 °C for 3 minutes. During the cool down, S gas was kept at a constant rate (2-5 sccm) to avoid material deterioration. Below 590°C, the flow is increased to 300 sccm for faster cooling down. $WS_2$ growth was the same as above but the growth temperature was set at 800°C and the sulfur source was melted (introduced) only after 760°C.

**Transfer of 2D materials and their heterostructures**. The transfer process of monolayer TMDs is shown in the Supporting Information Fig. S1. As-grown CVD $MoS_2$ or $WS_2$ monolayer on a $SiO_2$/Si substrate was cut into long narrow strips, attached to a PDMS film, and baked at 80 °C for 1 hr to ensure a good adhesion between the PDMS and the substrate. Then it was floated on 1M KOH solution until the substrate was etched off. The cutting of long-strip samples could decrease etching time, which usually took only 1-2 hrs. The left $MoS_2$/PDMS or $WS_2$/PDMS was rinsed in DI water for several times and then dried naturally. Afterwards, $MoS_2$/PDMS or $WS_2$/PDMS was attached onto a holey substrate, and kept there for ~ 2 hrs. PDMS film was then peeled off slowly, leaving $MoS_2$ or $WS_2$ monolayers on the holey substrate. $MoS_2$/$WS_2$ heterostructure was fabricated by stamping firstly $WS_2$ and then $MoS_2$, with a control of overlapping both monolayer parts. To make a $MoS_2$/graphene heterostructure, a graphene monolayer was firstly exfoliated onto a holey substrate, and then CVD $MoS_2$ monolayer was located and stamped onto the graphene monolayer.

  PDMS residues may reduce the interlayer interaction in the heterostructures. In our experiments, the KOH etching time and the PDMS attaching time on the target substrate are optimized to reduce the PDMS residues as much as possible. After the transfer, small PDMS particles were found occasionally to leave on some monolayers in AFM images.  However, these particles distributed sparsely and occupied only a very small fraction of the whole 2D area. Therefore, the reduction of the interlayer interaction due to the PDMS residues was low and negligible, and did not introduce large errors in the measurements.

**Measurement of modulus.** Holey substrates were fabricated by a deep UV photolithography followed by a deep reactive ion etching. The size of circular holes is diameter=1.1 μm and depth=0.3 μm. To measure elastic properties of 2D layers, a force (200-450 nN) was applied at the center of the suspended circular membranes through an AFM (Veeco Multimode) tip. The tip is coated by diamond-like carbon and its diameter is ~ 20 nm (μmash, HQ: NSC15). The indentation depth, δ, was determined by $\Delta z_{piezo} - \Delta z_{tip}$, where $\Delta z_{piezo}$ is the displacement of the scanning piezo-tube of AFM and $\Delta z_{tip}$ is the deflection of the AFM tip. The spring constant of the tip is 43.8 N/m, which was calibrated by a reference cantilever. All of the samples in this work, including monolayers and heterostructures, were measured by the same AFM tip to avoid errors introduced by using different tips.

**Calculation of modulus.** In the GGA-PBE approximation, we used a $9 \times 9 \times 2$ Gamma-point centered k-point mesh to sample the Brillouin zone and a plane-wave cutoff of 500 eV. The single-layer structures were simulated by the supercell approach under the periodic boundary condition. The vacuum region was chosen as large as 12 Å to avoid the interaction between monolayers in neighboring cells. Upon ionic relaxation, all atomic positions were optimized and the maximum force on each atom was less than 0.01 eV/Å. The unit cell and x-y directions were shown in Fig. S8. The unit cell parameters we obtained were nearly the same for $MoS_2$ (a= 3.190 nm) and $WS_2$ (a= 3.191 nm). Various strains along x and y directions were applied and the total energies at each points were calculated. The data points were fitted with the formula $E_s = a_1\varepsilon_x^2 + a_2\varepsilon_y^2 + a_3\varepsilon_x\varepsilon_y$ by the least-squares method, where $E_s$ is the energy deviation of the strained system from the equilibrium unstrained system, and $\varepsilon_x$ and $\varepsilon_y$ are the strain applied in x and y directions, respectively. Due to the isotropy of honeycomb lattice, $a_1$ equals $a_2$. The fitting coefficients determined the elastic modulus $Y$ and Poisson's ratio $\nu$, following

$Y = (2a_1 - a_3^2 / 2a_1) / A_0$ and $\nu = a_3 / 2a_1$, where $A_0$ is the area of the unit cell.[36, 37]

## ASSOCIATED CONTENT

**Supporting Information**. Transfer process of monolayer $MoS_2$ and $WS_2$, $MoS_2$ monolayers transferred on a holey substrate, photoluminescence (PL) and Raman spectra of $MoS_2$, as-grown and as-transferred $WS_2$, model for calculation of elastic properties of $MoS_2$ and $WS_2$, $MoS_2/WS_2$ heterostructure, exfoliated $MoS_2$ bilayer, $MoS_2$/graphene heterostructure.

## AUTHOR INFORMATION

**Corresponding Author**

*wuj@berkeley.edu

**Notes**

The authors declare no competing financial interest.

## ACKNOWLEDGMENT

This work was supported by the Office of Science, Office of Basic Energy Sciences, of the U.S. Department of Energy under Contract No. DE-AC02-05CH11231.The AFM characterization was partially supported by the NSF Center for Energy Efficient Electronics Science (NSF Award No. ECCS-0939514).Theoretical work at the Molecular Foundry was supported by the Office of Science, Office of Basic Energy Sciences of the U.S. Department of Energy under Contract No. DE-AC02-05CH11231. We thank Cong Liu for assistance in fabricating the holey substrates.

REFERENCES

1. Wang, Q. H.; Kalantar-Zadeh, K.; Kis, A.; Coleman, J. N.; Strano, M. S. *Nat. Nanotechnol.* **2012,** 7, (11), 699-712.
2. Butler, S. Z.; Hollen, S. M.; Cao, L. Y.; Cui, Y.; Gupta, J. A.; Gutierrez, H. R.; Heinz, T. F.; Hong, S. S.; Huang, J. X.; Ismach, A. F.; Johnston-Halperin, E.; Kuno, M.; Plashnitsa, V. V.; Robinson, R. D.; Ruoff, R. S.; Salahuddin, S.; Shan, J.; Shi, L.; Spencer, M. G.; Terrones, M.; Windl, W.; Goldberger, J. E. *Acs Nano* **2013,** 7, (4), 2898-2926.
3. Mak, K. F.; Lee, C.; Hone, J.; Shan, J.; Heinz, T. F. *Phys. Rev. Lett.* **2010,** 105, (13), 4.
4. Splendiani, A.; Sun, L.; Zhang, Y. B.; Li, T. S.; Kim, J.; Chim, C. Y.; Galli, G.; Wang, F. *Nano Letters* **2010,** 10, (4), 1271-1275.
5. Zhang, Y.; Chang, T. R.; Zhou, B.; Cui, Y. T.; Yan, H.; Liu, Z. K.; Schmitt, F.; Lee, J.; Moore, R.; Chen, Y. L.; Lin, H.; Jeng, H. T.; Mo, S. K.; Hussain, Z.; Bansil, A.; Shen, Z. X. *Nat. Nanotechnol.* **2014,** 9, (2), 111-115.
6. Jin, W.; Yeh, P.-C.; Zaki, N.; Zhang, D.; Sadowski, J. T.; Al-Mahboob, A.; van der Zande, A. M.; Chenet, D. A.; Dadap, J. I.; Herman, I. P.; Sutter, P.; Hone, J.; Osgood, R. M. *Phys. Rev. Lett.* **2013,** 111, (10).
7. Lopez-Sanchez, O.; Lembke, D.; Kayci, M.; Radenovic, A.; Kis, A. *Nat. Nanotechnol.* **2013,** 8, (7), 497-501.
8. Radisavljevic, B.; Radenovic, A.; Brivio, J.; Giacometti, V.; Kis, A. *Nat Nanotechnol* **2011,** 6, (3), 147-50.
9. Yoon, Y.; Ganapathi, K.; Salahuddin, S. *Nano Letters* **2011,** 11, (9), 3768-3773.
10. Lee, Y. H.; Zhang, X. Q.; Zhang, W. J.; Chang, M. T.; Lin, C. T.; Chang, K. D.; Yu, Y. C.; Wang, J. T. W.; Chang, C. S.; Li, L. J.; Lin, T. W. *Advanced materials* **2012,** 24, (17), 2320-2325.
11. Liu, K. K.; Zhang, W. J.; Lee, Y. H.; Lin, Y. C.; Chang, M. T.; Su, C.; Chang, C. S.; Li, H.; Shi, Y. M.; Zhang, H.; Lai, C. S.; Li, L. J. *Nano Letters* **2012,** 12, (3), 1538-1544.
12. Najmaei, S.; Liu, Z.; Zhou, W.; Zou, X. L.; Shi, G.; Lei, S. D.; Yakobson, B. I.; Idrobo, J. C.; Ajayan, P. M.; Lou, J. *Nat. Mater.* **2013,** 12, (8), 754-759.
13. Shi, Y. M.; Zhou, W.; Lu, A. Y.; Fang, W. J.; Lee, Y. H.; Hsu, A. L.; Kim, S. M.; Kim, K. K.; Yang, H. Y.; Li, L. J.; Idrobo, J. C.; Kong, J. *Nano Letters* **2012,** 12, (6), 2784-2791.
14. van der Zande, A. M.; Huang, P. Y.; Chenet, D. A.; Berkelbach, T. C.; You, Y. M.; Lee, G. H.; Heinz, T. F.; Reichman, D. R.; Muller, D. A.; Hone, J. C. *Nat. Mater.* **2013,** 12, (6), 554-561.
15. Geim, A. K.; Grigorieva, I. V. *Nature* **2013,** 499, (7459), 419-25.
16. Yu, W. J.; Li, Z.; Zhou, H. L.; Chen, Y.; Wang, Y.; Huang, Y.; Duan, X. F. *Nat. Mater.* **2013,** 12, (3), 246-252.

17. Yu, W. J.; Liu, Y.; Zhou, H. L.; Yin, A. X.; Li, Z.; Huang, Y.; Duan, X. F. *Nat. Nanotechnol.* **2013,** 8, (12), 952-958.
18. Britnell, L.; Ribeiro, R. M.; Eckmann, A.; Jalil, R.; Belle, B. D.; Mishchenko, A.; Kim, Y. J.; Gorbachev, R. V.; Georgiou, T.; Morozov, S. V.; Grigorenko, A. N.; Geim, A. K.; Casiraghi, C.; Neto, A. H. C.; Novoselov, K. S. *Science* **2013,** 340, (6138), 1311-1314.
19. Tongay, S.; Fan, W.; Kang, J.; Park, J.; Koldemir, U.; Suh, J.; Narang, D. S.; Liu, K.; Ji, J.; Li, J.; Sinclair, R.; Wu, J. *Nano Lett* **2014,** 14, (6), 3185-90.
20. Rogers, J. A.; Someya, T.; Huang, Y. G. *Science* **2010,** 327, (5973), 1603-1607.
21. He, Q. Y.; Zeng, Z. Y.; Yin, Z. Y.; Li, H.; Wu, S. X.; Huang, X.; Zhang, H. *Small* **2012,** 8, (19), 2994-2999.
22. Pu, J.; Yomogida, Y.; Liu, K. K.; Li, L. J.; Iwasa, Y.; Takenobu, T. *Nano Letters* **2012,** 12, (8), 4013-4017.
23. Chang, H. Y.; Yang, S. X.; Lee, J. H.; Tao, L.; Hwang, W. S.; Jena, D.; Lu, N. S.; Akinwande, D. *Acs Nano* **2013,** 7, (6), 5446-5452.
24. Salvatore, G. A.; Munzenrieder, N.; Barraud, C.; Petti, L.; Zysset, C.; Buthe, L.; Ensslin, K.; Troster, G. *Acs Nano* **2013,** 7, (10), 8809-8815.
25. Bertolazzi, S.; Brivio, J.; Kis, A. *ACS Nano* **2012,** 5, (12), 9703-9.
26. Castellanos-Gomez, A.; Poot, M.; Steele, G. A.; van der Zant, H. S.; Agrait, N.; Rubio-Bollinger, G. *Advanced materials* **2012,** 24, (6), 772-5.
27. Cooper, R.; Lee, C.; Marianetti, C.; Wei, X.; Hone, J.; Kysar, J. *Physical Review B* **2013,** 87, (3).
28. Lee, C.; Yan, H.; Brus, L. E.; Heinz, T. F.; Hone, J.; Ryu, S. *Acs Nano* **2010,** 4, (5), 2695-2700.
29. Li, H.; Zhang, Q.; Yap, C. C. R.; Tay, B. K.; Edwin, T. H. T.; Olivier, A.; Baillargeat, D. *Adv. Funct. Mater.* **2012,** 22, (7), 1385-1390.
30. Conley, H. J.; Wang, B.; Ziegler, J. I.; Haglund, R. F.; Pantelides, S. T.; Bolotin, K. I. *Nano Letters* **2013,** 13, (8), 3626-3630.
31. Tongay, S.; Zhou, J.; Ataca, C.; Liu, J.; Kang, J. S.; Matthews, T. S.; You, L.; Li, J.; Grossman, J. C.; Wu, J. *Nano Letters* **2013,** 13, (6), 2831-2836.
32. Lee, G. H.; Cooper, R. C.; An, S. J.; Lee, S.; van der Zande, A.; Petrone, N.; Hammerberg, A. G.; Lee, C.; Crawford, B.; Oliver, W.; Kysar, J. W.; Hone, J. *Science* **2013,** 340, (6136), 1073-6.
33. Wan, K.-T.; S., G.; Dillard, D. A. *Thin Solid Films* **2003,** 425, 150-162.
34. Komaragiri, U.; Begley, M. R.; Simmonds, J. G. *Journal of Applied Mechanics* **2005,** 72, (2), 203.
35. Lee, C.; Wei, X.; Kysar, J. W.; Hone, J. *Science* **2008,** 321, (5887), 385-8.
36. Topsakal, M.; Cahangirov, S.; Ciraci, S. *Applied Physics Letters* **2010,** 96, (9), 091912.
37. Yue, Q.; Kang, J.; Shao, Z.; Zhang, X.; Chang, S.; Wang, G.; Qin, S.; Li, J. *Physics Letters A* **2012,** 376, (12-13), 1166-1170.
38. Feldman, J. L. *J. Phys. Chem. Solids* **1976,** 37, 1141-4.
39. Filippi, C.; Singh, D.; Umrigar, C. *Physical Review B* **1994,** 50, (20), 14947-14951.
40. Liu, Z.; Yang, J.; Grey, F.; Liu, J. Z.; Liu, Y.; Wang, Y.; Yang, Y.; Cheng, Y.; Zheng, Q. *Phys. Rev. Lett.* **2012,** 108, (20).
41. Kis, A.; Jensen, K.; Aloni, S.; Mickelson, W.; Zettl, A. *Phys. Rev. Lett.* **2006,** 97, (2).
42. Nigues, A.; Siria, A.; Vincent, P.; Poncharal, P.; Bocquet, L. *Nat Mater* **2014,** 13, (7), 688-93.

43. Kaplan-Ashiri, I.; Cohen, S. R.; Gartsman, K.; Ivanovskaya, V.; Heine, T.; Seifert, G.; Wiesel, I.; Wagner, H. D.; Tenne, R. *Proceedings of the National Academy of Sciences of the United States of America* **2006,** 103, (3), 523-8.

Supporting Information for

# Elastic Properties of Chemical-Vapor-Deposited Monolayer MoS$_2$, WS$_2$, and Their Bilayer Heterostructures

*Kai Liu,*[†,‡] *Qimin Yan,*[#,∥] *Michelle Chen,*[†] *Wen Fan,*[†,□] *Yinghui Sun,*[∥] *Joonki Suh,*[†] *Deyi Fu,*[†] *Sangwook Lee,*[†] *Jian Zhou,*[†] *Sefaattin Tongay,*[†] *Jie Ji,*[□] *Jeffrey B. Neaton,*[#,∥] *Junqiao Wu*[†,‡,*]

[†] Department of Materials Science and Engineering, University of California, Berkeley, California 94720, United States

[‡] Materials Sciences Division, Lawrence Berkeley National Laboratory, Berkeley, California 94720, United States

[#] Molecular Foundry, Lawrence Berkeley National Laboratory, Berkeley, California 94720, United States

[∥] Department of Physics, University of California, Berkeley, California 94720, United States

[□] Department of Thermal Science and Energy Engineering, University of Science and Technology of China, Anhui 230027, China

* Corresponding author. Email: wuj@berkeley.edu

## 1. Transfer process of monolayer MoS$_2$ and WS$_2$

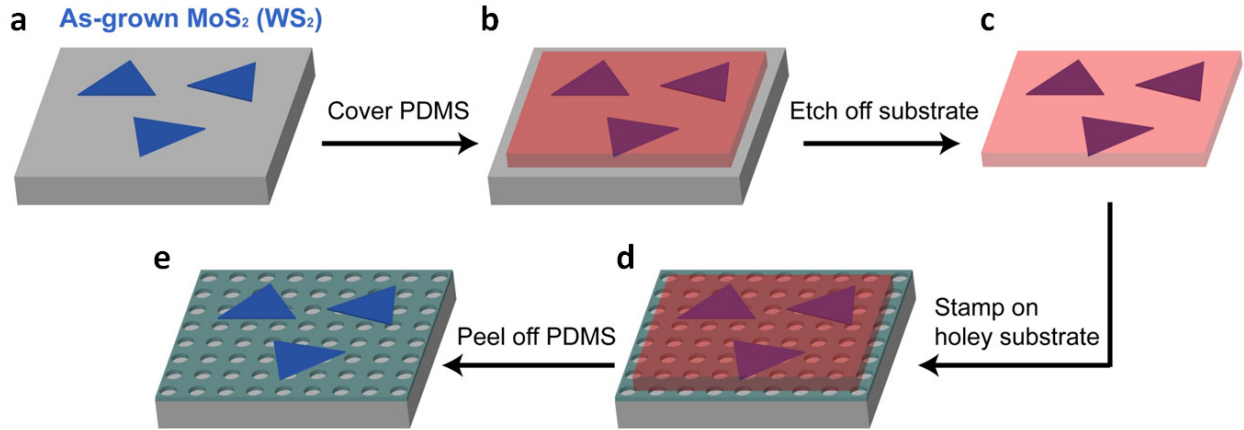

Figure S1. PDMS stamp of chemical-vapor-deposited 2D monolayers.

## 2. MoS$_2$ monolayers transferred on a holey substrate

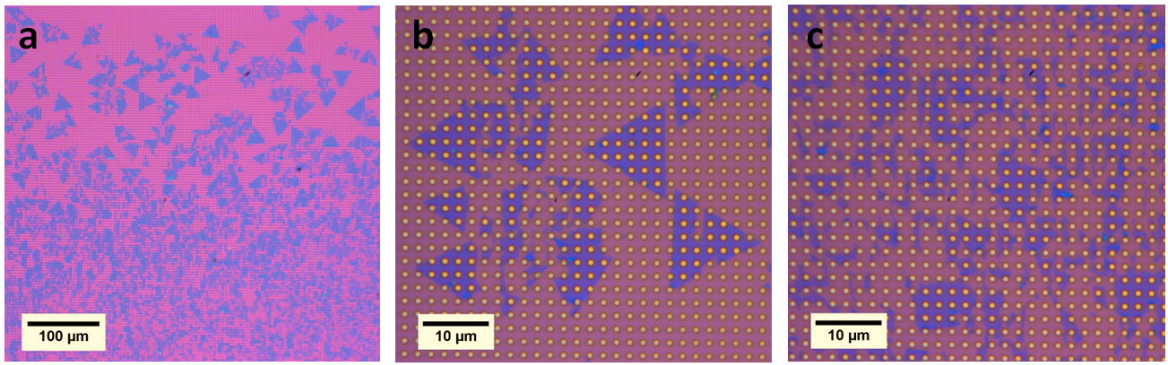

Figure S2. (a) Overall view of both isolated triangles and continuous part as grown on substrate. (b) Isolated triangle monolayers transferred onto the holey substrate. It shows most of triangles are transferred, and many of them are intact after the transfer. (c) Continuous monolayer part after transfer. The originally continuous film breaks into isolated pieces. The transfer yield is ~70 % in area for the triangle part and ~ 20 % for the continuous part. On a flat SiO$_2$/Si substrate, in contrast, almost 100% of triangle and continuous parts can be transferred.

## 3. Comparison of isolated and continuous parts of CVD MoS$_2$ monolayers

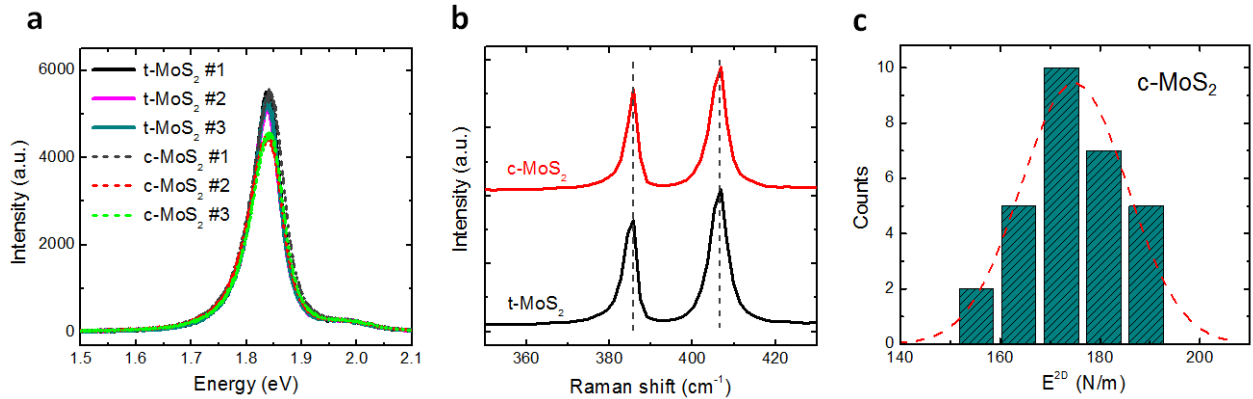

Figure S3. (a) Photoluminescence of triangle (t-MoS$_2$) and continuous (c-MoS$_2$) parts. Three different positions were probed for either part. (b) Raman spectra of both parts. (c) Statistical histogram of the 2D modulus of the continuous part of MoS$_2$. The very similar PL and Raman spectra, as well as the nearly identical modulus (171 ± 11 N/m for t-MoS$_2$ and 174 ± 10 N/m for c-MoS$_2$), suggest that these two monolayer parts have similar crystal quality.

## 4. As-grown and as-transferred WS$_2$

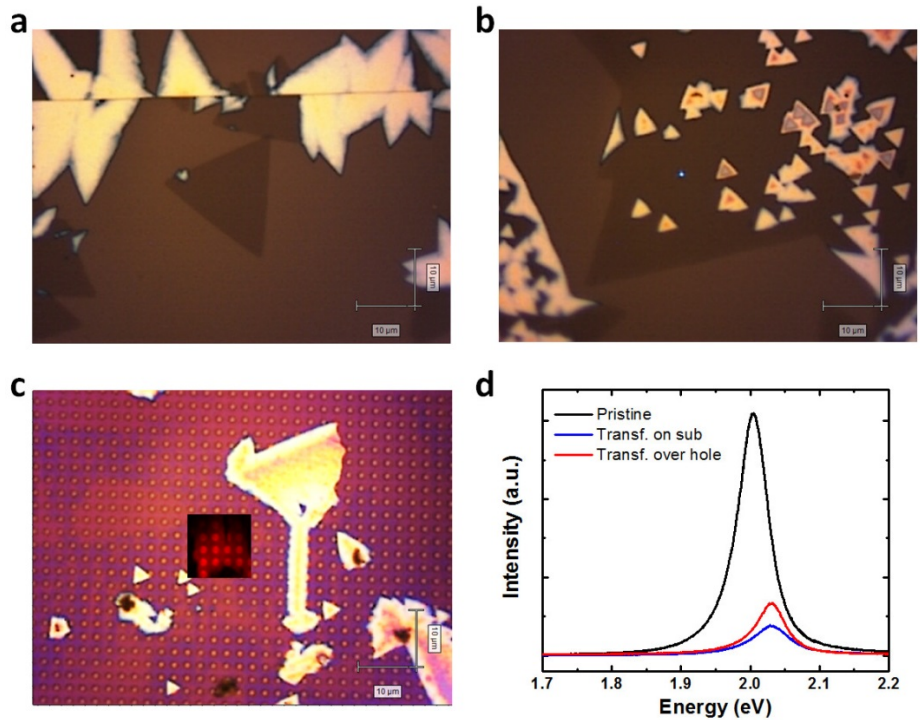

Figure S4. (a, b) Optical images of as-grown WS$_2$ sample. CVD WS$_2$ sample also has isolated triangles and relatively continuous part, but many thick WS$_2$ parts (bright color in images) form around the triangle and continuous parts, which is different from MoS$_2$ growth. (c) As-transferred WS$_2$ monolayer on a holey substrate. The inset shows PL mapping of the corresponding area. The suspended circular membranes present stronger PL intensities, which is the same as those of MoS$_2$. (d) PL spectra of as-grown and as-transferred WS$_2$. The as-grown WS$_2$ exhibits a PL peak at 2.00 eV originating from the A excitonic transition,[1, 2] while after transfer, this peak is blue shifted to 2.03 eV. This trend is similar to that of MoS$_2$ shown in Fig. 1g.

## 5. MoS$_2$/WS$_2$ heterostructure

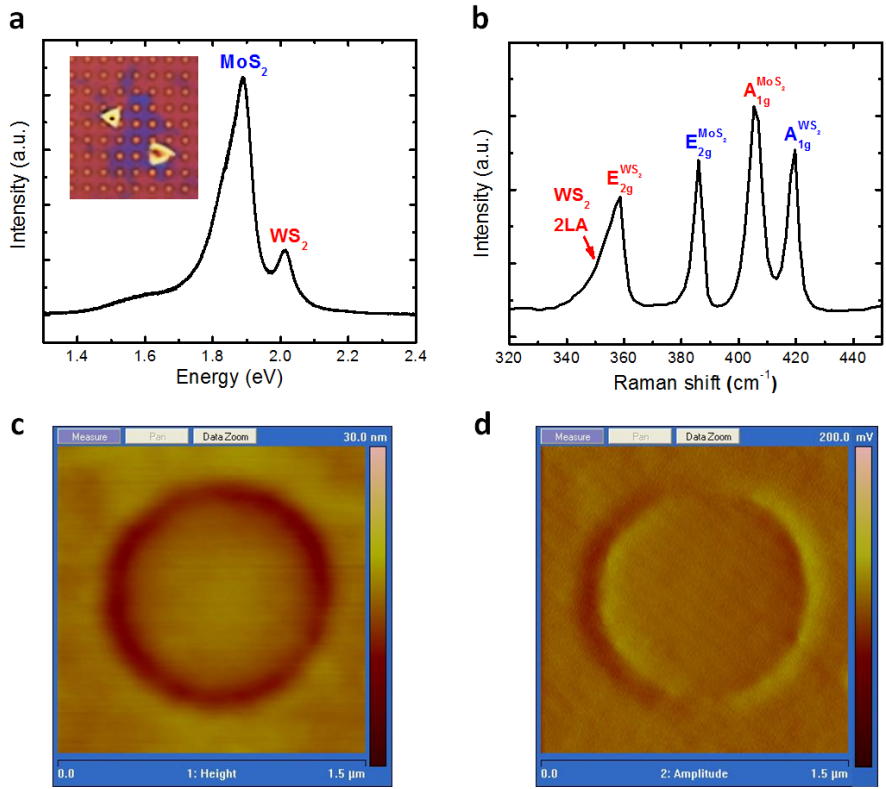

Figure S5. (a) PL spectrum of a suspended $MoS_2/WS_2$ hetero membrane in the area shown in the inset, demonstrating the overlap of PL peaks from direct band gap transitions of $MoS_2$ and $WS_2$. $MoS_2/WS_2$ heterostructure was fabricated by a two-step stamping transfer of firstly $WS_2$ and then $MoS_2$. We found that the second-step stamp peeled off some $WS_2$ monolayers that are previously transferred on the holey substrate, but also overlapped $MoS_2$ to $WS_2$ in some areas. (b) Raman spectrum of as-fabricated $MoS_2/WS_2$, also showing the combination of both Raman features.[3,4] (c and d) AFM topology and amplitude images of a $MoS_2/WS_2$ hetero membrane, which do not show any bubbles or wrinkles on the substrate or over the hole.

## 6. Exfoliated MoS$_2$ bilayer

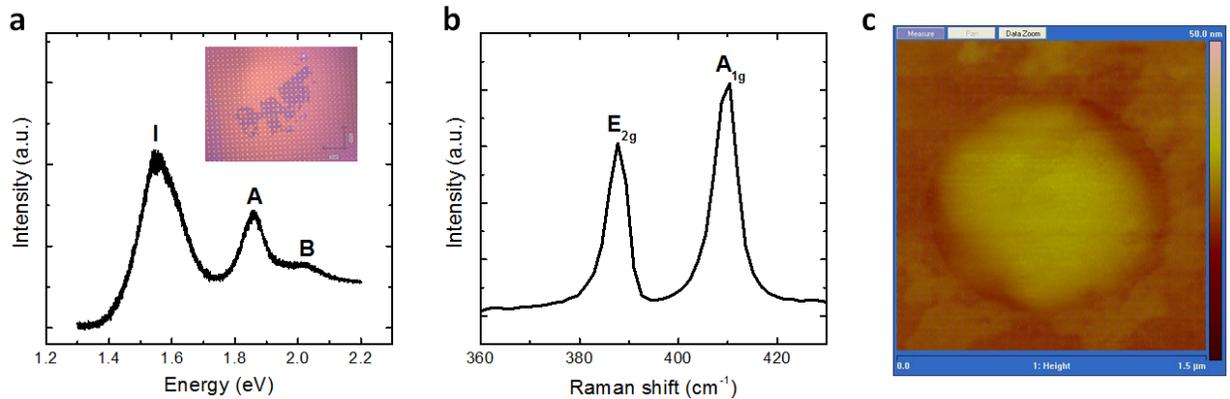

Figure S6. (a) PL spectrum of an exfoliated $MoS_2$ bilayer shown in the inset. The spectrum presents peaks of A and B excitonic transitions (centered at 1.86 eV and 2.03 eV, respectively), and indirect band gap transition, I (centered at 1.55 eV). (b) Raman spectrum of $MoS_2$ bilayer. The interval between $E_{2g}$ and $A_{1g}$ is 22 cm$^{-1}$. Both the PL and the Raman spectra match that of $MoS_2$ bilayer reported in literature.[3,5] (c) AFM topology image of a $MoS_2$ bilayer membrane.

## 7. MoS$_2$/graphene heterostructure

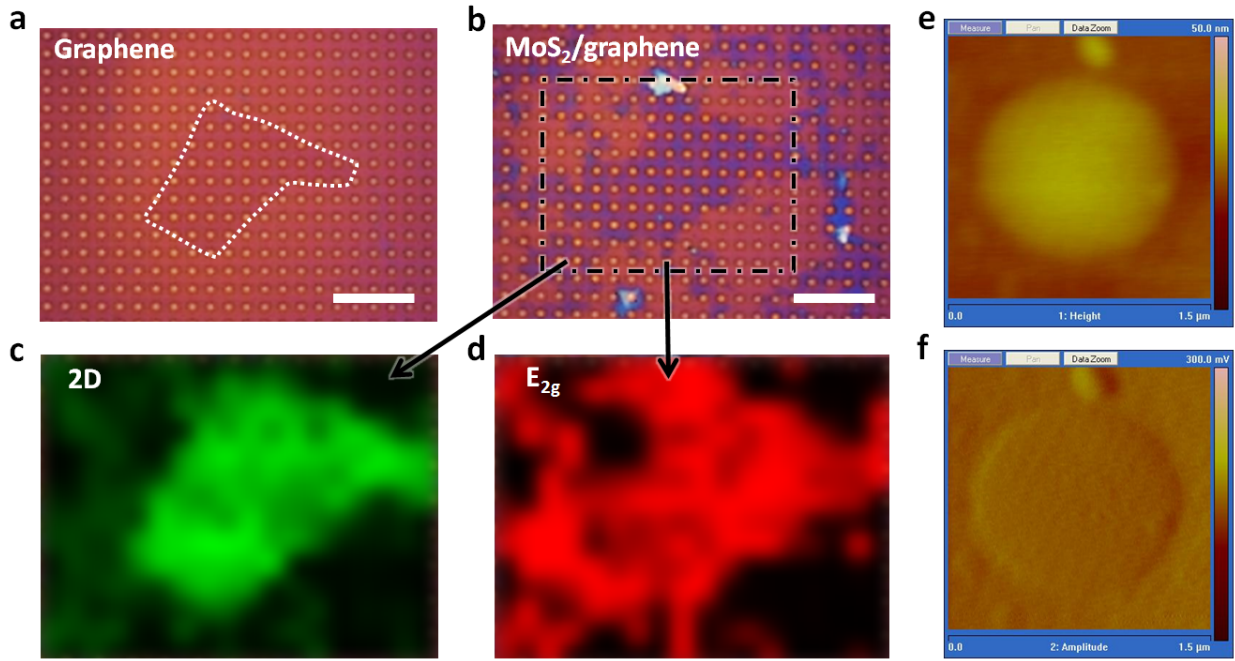

Figure S7. (a) As-exfoliated graphene on a holey substrate. The white dotted line shows the edge of the graphene monolayer. (b) Transfer of CVD MoS$_2$ by PDMS stamp, overlapping MoS$_2$ continuous monolayer on graphene and forming MoS$_2$/graphene heterostructure. The area labeled by black dash-dot line is mapped in Raman spectra. (c) Raman mapping of 2D peak (2692 cm$^{-1}$), showing the area of graphene. (d) Raman mapping of E$_{2g}$ peak (387 cm$^{-1}$), showing the area of MoS$_2$. (e and f) AFM topology and amplitude images of a MoS$_2$/graphene hetero membrane. No evidence of bubbles or wrinkles exists on the substrate or over the hole.

## 8. Model for the calculation of elastic properties of MoS$_2$ and WS$_2$

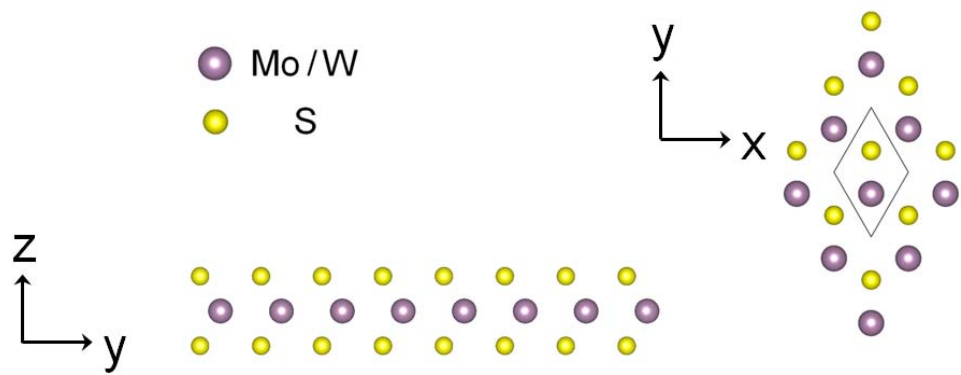

Figure S8. Top (right) and side view (bottom) of monolayer $MoS_2$ or $WS_2$. The unit cell used in our simulations is marked by the black diamond box. Two in-plane directions (x and y) are represented by the arrows.

**References**

1. Zhao, W. J.; Ghorannevis, Z.; Chu, L. Q.; Toh, M. L.; Kloc, C.; Tan, P. H.; Eda, G. *Acs Nano* **2013,** 7, (1), 791-797.
2. Gutierrez, H. R.; Perea-Lopez, N.; Elias, A. L.; Berkdemir, A.; Wang, B.; Lv, R.; Lopez-Urias, F.; Crespi, V. H.; Terrones, H.; Terrones, M. *Nano Letters* **2013,** 13, (8), 3447-3454.
3. Lee, C.; Yan, H.; Brus, L. E.; Heinz, T. F.; Hone, J.; Ryu, S. *Acs Nano* **2010,** 4, (5), 2695-2700.
4. Berkdemir, A.; Gutiérrez, H. R.; Botello-Méndez, A. R.; Perea-López, N.; Elías, A. L.; Chia, C.-I.; Wang, B.; Crespi, V. H.; López-Urías, F.; Charlier, J.-C.; Terrones, H.; Terrones, M. *Scientific Reports* **2013,** 3.
5. Mak, K. F.; Lee, C.; Hone, J.; Shan, J.; Heinz, T. F. *Phys. Rev. Lett.* **2010,** 105, (13), 4.